# Full-field modeling of heat transfer in asteroid regolith: Radiative thermal conductivity of polydisperse particulates

[1,2]Andrew J. Ryan, [3]Daniel Pino Muñoz, [3]Marc Bernacki, [1]Marco Delbo

[1]Université Côte d'Azur, Observatoire de la Côte d'Azur, CNRS, Laboratoire Lagrange, Nice, France; [2]Now at: Lunar and Planetary Laboratory, University of Arizona, Tucson, AZ, USA. [3]Mines ParisTech, PSL Research University, CEMEF - Centre de mise en forme des matériaux, Sophia Antipolis Cedex, France.

Corresponding author: Andrew Ryan (ajryan@orex.lpl.arizona.edu)

**Key Points:**
- A new finite element model for analyzing the solid and radiative conductivity of particulate regoliths is presented.
- Non-isothermality in particles with low material thermal conductivity or large sizes can markedly lower the radiative thermal conductivity
- Particulate size mixtures are shown to have a radiative thermal conductivity that is equivalent to the Sauter mean particle diameter

**Abstract**

Characterizing the surface material of an asteroid is important for understanding its geology and for informing mission decisions, such as the selection of a sample site. Diurnal surface temperature amplitudes are directly related to the thermal properties of the materials on the surface. We describe a numerical model for studying the thermal conductivity of particulate regolith in vacuum. Heat diffusion and surface-to-surface radiation calculations are performed using the finite element (FE) method in three-dimensional meshed geometries of randomly packed spherical particles. We validate the model for test cases where the total solid and radiative conductivity values of particulates with monodisperse particle size frequency distributions (SFDs) are determined at steady-state thermal conditions. Then, we use the model to study the bulk radiative thermal conductivity of particulates with polydisperse, cumulative power-law particle SFDs. We show that for each polydisperse particulate geometry tested, there is a corresponding monodisperse geometry with some *effective* particle diameter that has an identical radiative thermal conductivity. These effective diameters are found to correspond very well to the Sauter mean particle diameter, which is essentially the surface-area-weighted mean. Next, we show that the thermal conductivity of the particle material can have an important effect on the radiative component of the thermal conductivity of particulates, especially if the particle material conductivity is very low or the spheres are relatively large, owing to non-isothermality in each particle. We provide an empirical correlation to predict the effects of non-isothermality on radiative thermal conductivity in both monodisperse and polydisperse particulates.

**Plain Language Summary**

The thermal conductivity of asteroid regolith is related to the properties of the particulate assemblage (e.g. size distribution). Spacecraft missions that measure the surface temperature of asteroids, like OSIRIS-REx at asteroid Bennu, can take advantage of this by relating the observed temperatures to the physical properties of the regolith. We present a 3D model for studying the thermal conductivity of regolith, where heat flow is simulated in randomly packed spheres. We found that for cases where the particle sizes are monodisperse, our model reproduces the thermal conductivity values predicted by simpler theoretical models. However, this is only true if the particles themselves are made of a material that itself has relatively high thermal conductivity, which may not be the case for the regolith on Bennu. We determined the values for a correction factor to account for these cases. Neglecting it could cause one to appreciably underestimate particle sizes on asteroid surfaces, which could pose a risk for sample collection. Finally, we found that regoliths with particle size mixtures can have radiative thermal conductivities that are identical to monodisperse regoliths. We found that the surface-area-weighted mean particle size of the mixed regoliths is representative of the bulk radiative thermal conductivity.

**1 Introduction**

Asteroids and airless bodies, such as the Moon, are commonly blanketed in loose and unconsolidated material known as regolith, which preserves a record of the geophysical history of the body (Murdoch et al., 2015). The particle size, maturity, and depth of the regolith layer are thought to vary within the asteroid populations depending on the size, composition, shape, and age of these bodies (Murdoch et al., 2015; and references therein). Regolith physical properties also place important constraints on mission planning, particularly for safety during sample acquisition events. NASA's Origins, Spectral Interpretation, Resource Identification, and Security–Regolith Explorer (OSIRIS-REx, Lauretta et al., 2019) and JAXA's Hayabusa2 (Watanabe et al., 2019) missions are currently studying the surface of carbonaceous asteroids (101955) Bennu and (162173) Ryugu, respectively. The prime objective for both missions is to collect regolith samples and return them to Earth. Infrared emission measurements of the asteroid surface by OSIRIS-REx's Thermal Emission Spectrometer (OTES, Christensen et al., 2018) and Visible and InfraRed Spectrometer (OVIRS, Reuter et al., 2018), and by Hayabusa2's Thermal Infrared Imager (TIR, Okada et al., 2017), are currently being used to infer the physical properties of the regolith, such as particle size distribution, porosity, and composition. This is possible because the amplitude of diurnal temperature cycling of the regolith is directly related to the thermal inertia of the regolith, which in turn is controlled by the physical characteristics of the regolith.

OSIRIS-REx and Hayabusa2 recently revealed that asteroids Bennu and Ryugu are both covered in abundant large boulders with very little fine regolith surface cover, which increases the challenge of sample site selection (Lauretta et al., 2019; DellaGiustina and Emery et al., 2019; Walsh et al., 2019; Sugita et al., 2019; Watanabe et al., 2019). Preliminary infrared telescopic observations of these asteroids were interpreted to represent regolith with particle diameters ranging from a few millimeters to about a centimeter for Bennu (Emery et al., 2014; Müller et al., 2017). However, the images of the surfaces of Bennu and Ryugu from both space missions show that this is not case. The mismatch between regolith size predictions and surface imaging observations could be due to unexpectedly low thermal conductivity and density of the solid material composing the boulders of these asteroids (Grott et al., 2019), limitations in the models used to interpret thermal infrared observations in term of regolith particle sizes, or a combination of both. In this work, we focus on the latter and seek to improve existing thermophysical models for asteroid regoliths.

Several theoretical models exist that predict the thermal conductivity of monodisperse particulate media as a function of particle size, packing structure, and physical parameters of the particles themselves, such as their thermal conductivity and emissivity (e.g., van Antwerpen et al., 2012; Gundlach & Blum, 2012, 2013; Sakatani et al., 2017). However, no such models have sought to describe polydisperse particulates, despite the prevalence of particle size mixtures in nature (except for where there are advanced natural sorting mechanisms in place). Furthermore, only some models (e.g., van Antwerpen et al., 2012) have sought to incorporate particulate non-isothermality effects, which could be important for regolith on Ryugu and Bennu, particularly if the geologic materials have unusually low densities and thermal conductivities.

Experimental studies have provided invaluable data necessary to calibrate theoretical models for particulate thermal conductivity, including some preliminary analyses of polydisperse samples (Sakatani et al., 2017, 2018; Ryan, 2018;). However, such experiments are known to be difficult to perform and can produce results with uncertainties that are difficult to quantify, e.g. due to factors that are difficult to control or know, such as changes in particulate porosity in the vicinity of heat sources/sinks (Presley & Christensen, 1997; Ryan, 2018). This limits the number of experiments that can reasonably be performed and the utility of the results. Furthermore, there is often uncertainty in the physical aspects of the samples being studied, such as the nature of the packing structure, the magnitude of contact deformation between particles, and relevant physical parameters of the particles themselves, such as their thermal conductivity and emissivity, that can complicate the process of relating experimental results to theoretical models. This problem leads us to the methods presented herein, where advanced numerical methods are used to create a realistic, though idealized, simulation of a thermal conductivity experiment in which physical parameters related to the "experimental" samples are well controlled and thus known with certainty.

We present a thermophysical model for particulates in which a bed of spherical particles is fully rendered in a three-dimensional finite element (FE) mesh framework. This type of model enables rapid and well-controlled investigations of the fundamental parameters that affect particulate thermal conductivity. In this work, we present validation cases for solid and radiative thermal conduction through ordered and randomly packed monodisperse particulates. Then, we present the findings of a study of the radiative thermal conductivity properties in mixtures that follow power-law particle size frequency distributions (SFDs) commonly observed in boulders and coarse regolith on asteroid surfaces (Bierhaus et al., 2018). Radiative thermal conductivity is selectively studied here because its magnitude tends to greatly exceed that of conduction through the solids for coarse particulates (> a few mm diameter, as explained in Section 2: Background and shown in Figure 1) at temperatures relevant to the study of near-Earth asteroids. Given the likely abundance of coarse materials on Bennu and Ryugu and the need to identify particle size thresholds below ~2 cm for sampling, the study of radiative thermal conductivity in coarse particulates mixtures is of high priority. We pay particular attention in these analyses to the effects of particle non-isothermality, given the initial findings by OSIRIS-REx and Hayabusa2 of geologic materials with potentially low thermal conductivity on Bennu and Ryugu (DellaGiustina and Emery et al., 2019; Sugita et al., 2019; Grott et al., 2019).

## 2 Background

Two modes of heat transfer exist in particulate media in a vacuum — radiation between particles and conduction through the solids at the contacts between particles (Wesselink, 1948; Watson, 1964; Wechsler et al., 1972; van Antwerpen et al., 2010; De Beer et al., 2018). These are typically represented in terms of their contributions to the bulk thermal conductivity of the regolith, where conductivity due to radiative heat transfer is denoted by $k_r$ and conductivity due to the contacts between the particles is denoted by $k_s$. In the presence of an atmosphere, a third heat conduction term must be considered to account to heat transfer through pore-filling gas. The gas thermal conductivity, $k_g$, tends to be the dominant heat transfer mechanism on Mars (Presley & Christensen, 1997). Although gas conduction will not be considered in this present work, a proper treatment of thermal conductivity due to radiation and particle contacts is still useful for improving models of regolith thermal conductivity on Mars, Earth, or other bodies with appreciable atmospheres.

For large, opaque particles, heat transfer by radiation is governed by the Stefan-Boltzmann law, which is influenced by the temperature, emissivity, and distance between radiating surfaces. Most theoretical formulations for radiative heat transfer in particulates begin by approximating the particles as a series of parallel, radiating plates separated by some representative distance for radiative transfer. The radiative thermal conductivity ($k_r$) of such a configuration is proportional to the cube of the local mean temperature (e.g., Wesselink, 1948). The representative distance for radiative transfer is often related to the size of the pore space between particles, which in turn is related to porosity and particle diameter, as is the case in the model for regolith thermal conductivity by Sakatani et al. (2017).

The regolith thermal conductivity model by Gundlach and Blum (2012; 2013) uses the mean free path of the photon as the representative distance for radiative transfer. This formulation is based on theory for very small particles (<100 µm diameter), where particle transparency becomes important (Merrill, 1969). As such, it should be used with caution for larger particulates, particularly given that Gundlach and Blum's model predicts a much stronger dependence between radiative thermal conductivity and porosity than most other theoretical models (e.g., Schotte, 1960; Watson, 1964; Breitbach & Barthels, 1980; van Antwerpen et al., 2012; Sakatani et al., 2017).

A more advanced formulation by van Antwerpen et al. (2012) separately describes the radiative contribution of immediately adjacent spheres (short-range radiation) and nonadjacent spheres that still have some direct line of sight (long-range radiation). They incorporate approximations to describe the view factors—i.e., the proportion of radiation emitted from one surface that reaches another—between spheres for both of their radiative conductivity formulations. Furthermore, they demonstrate that the thermal conductivity of the solid material that composes the spheres can influence the radiative thermal conductivity, building upon work by Breitbach and Barthels (1980) and Singh and Kaviany (1994). In such cases, thermal gradients are present within individual particles, with the consequence being that the radiative thermal conductivity would not increase in proportion to the cube of the temperature but rather at a lower rate. Van Antwerpen et al. (2012) introduce an updated description of a non-isothermal correction factor, $f_k$, that is related to a dimensionless parameter, $\Lambda_s$:

(Eq. 1)
$$\Lambda_s = \frac{k_m}{4D\sigma T^3}$$

where $k_m$ is the thermal conductivity of the material that constitutes the particles (the thermal conductivity of a single particle), $D$ is particle diameter, $\sigma$ is the Stefan-Boltzmann constant, and $T$ is the mean local temperature. If $1/\Lambda_s$ is greater than approximately 0.04, the particle non-isothermality can have a meaningful, measurable effect (>~1%) on radiative thermal conductivity. Van Antwerpen et al. (2012) provide correlation coefficients to relate $\Lambda_s$ to the correction factor, $f_k$ in the direction of heat flow:

(Eq. 2)
$$f_k = a_1 \tan^{-1}\left(a_2 \left(\frac{1}{\Lambda_s}\right)^{a_3}\right) + a_4$$

When emissivity is unity, $a_1 = -0.3966$, $a_2 = 0.7495$, $a_3 = 0.5738$, and $a_4 = 1.0484$. When $1/\Lambda_s < 0.01$, $f_k = 1$. These four correction factor coefficients were determined from numerical simulations of radiative heat transfer between two almost-touching hemispheres, implying that they are accurate for correcting short-range radiation non-isothermality. Van Antwerpen et al. (2012) use the non-isothermal correction factor for both short- and long-range radiative conductivity by simply multiplying each conductivity by $f_k$ but note that more work is needed to verify its validity for long-range radiation. Experimental and theoretical work describing particle thermal conductivity in the context of planetary regoliths has yet to incorporate non-isothermality in the radiative thermal conductivity, primarily owing to the normally small size of the particles under consideration (typically <1 mm) and the relatively high thermal conductivity of geologic materials (e.g., basalt: ~1–2.5 W m$^{-1}$ K$^{-1}$ (Desai et al., 1974)). Finally, it should be noted that even though the thermophysical properties of a planetary surface are determined in terms of thermal inertia, which is related to the square root of thermal conductivity, the effects of non-isothermality could still considerably affect the estimation of particle size on rubble pile asteroids like Bennu and Ryugu.

Having discussed the thermal conductivity of particulates due to radiation ($k_r$), we now move our attention to the thermal conduction through the particle-to-particle contacts. The bulk thermal conductivity of particulates due to these contacts ($k_s$) is directly related to but typically much smaller than the thermal conductivity of the individual particles themselves (denoted here by $k_m$ for *material* conductivity). Although the thermal conductivity of most common geologic materials is relatively high (~2 to 4 W m$^{-1}$ K$^{-1}$), the flow of heat from one particle to the next is severely hindered by the very small size of the contacts between particles. The size of these contacts and the average number of contacts between particles (the coordination number) thus strongly influence the solid thermal conductivity of particulates. To estimate the size of the contacts, Gundlach and Blum (2013) and Sakatani et al. (2017) use the model for interparticle forces due to adhesion and the resulting Hertzian deformation by Johnson, Kendall, & Roberts (1971) ("JKR theory") to predict the particle contact radii that are used in their models for the thermal conductivity of particulates. The contact radii $r_c$ are typically very small compared with the radii of the particles $R_p$, with values for the ratio $r_c/R_p$ in the range of 0.0005–0.005 (depending on values for particle elastic moduli, surface energy, and gravity). As such, the magnitude of thermal conductivity by means of heat transfer between the particles via their contacts, $k_s$, is typically several factors or even orders of magnitude smaller than the thermal conductivity of the particle material itself, $k_m$. Sakatani et al. (2017) have proposed that $k_s$ and $k_m$ are linearly related to each other by the ratio $r_c/R_p$ (Sakatani et al., 2017). Also, they and others (e.g. van Antwerpen et al., 2012) have propose that $k_s$ is linearly related to the mean coordination number. Several model correlations between coordination number and porosity have been proposed (see van Antwerpen et al. (2010) for a review), but more work is needed on this topic to resolve discrepancies between models and to determine the effects of particle sphericity and roughness.

Figure 1 illustrates the relative contributions of radiative thermal conductivity and solid conductivity to the total thermal conductivity of particulates on Bennu, as predicted by the particulate thermal conductivity model by Sakatani et al. (2017). The radiative thermal conductivity increases linearly with particle diameter, whereas solid thermal conductivity only changes modestly (or not at all, if adhesion is neglected). Therefore, the radiative thermal conductivity term comes to dominate for particles larger than a few millimeters in diameter.

All existing theoretical models for regolith thermal conductivity assume monodisperse particle SFDs (all particles have the same size). Some experimental work, however, has been conducted to determine whether these models could be used to approximate the thermal properties of particulates with polydisperse SFDs. Sakatani et al. (2018) measured the thermal properties of the JSC-1 lunar regolith simulant (maximum diameter <=~1 mm) with a log-normal cumulative particle mass distribution and found a thermal conductivity approximately equivalent to that of the volumetric median particle size (~100 μm). However, it remains to be determined whether this relationship is applicable to other particle size distributions and whether it holds for both radiative and solid conductivity terms.

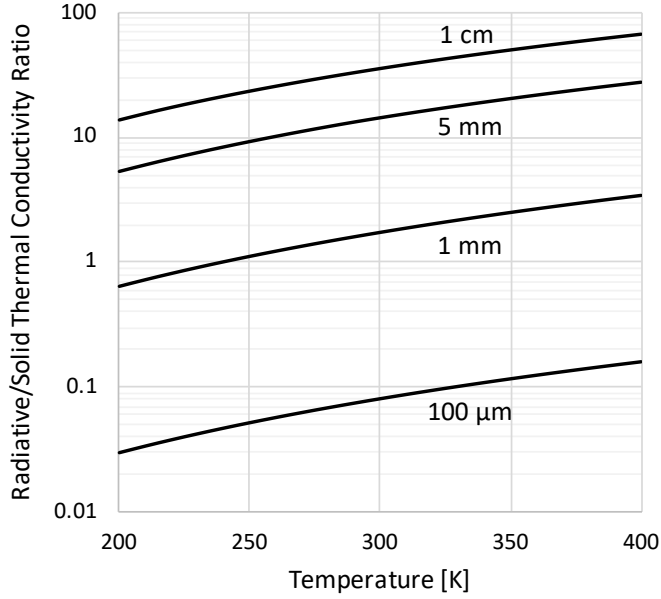

**Figure 1.** Illustrative example of the predicted ratio of radiative to solid thermal conductivity (i.e. $k_r/k_s$) in monodisperse particulates on Bennu as a function of particle diameter (indicated next to the lines). Values were calculated with the model from Sakatani et al. (2017), using parameters assumed for asteroid Bennu, where particulate material properties are similar to CM chondrite Cold Bokkeveld (Opeil et al., 2010), surface energy is 0.02 J m$^{-2}$, depth of burial is 1 cm, gravity is 75 μm s$^{-2}$, emissivity is 1.0, and porosity is 0.40. Van Antwerpen et al. (2012) and Gundlach and Blum (2013) models yield comparable predictions when the same or similar input parameters are used.

## 3 Methods and Model Validation

We study the thermal conductivity of particulates in a vacuum within a FE framework where constant heat flux is applied to one side of a three-dimensional, parallelepiped geometry of randomly packed spheres while a constant temperature boundary condition is applied to the opposite side (Figures 2 & 3). Once the model temperatures reach steady state, the thermal conductivity of the particulates can be calculated from the temperature differential between the two opposing geometry boundaries where the flux and fixed temperature boundary conditions have been applied:

(Eq. 3)
$$k = q \frac{\Delta x}{\Delta T}$$

Or in the case where the contribution of the two plates is removed:

(Eq. 4)
$$k = \frac{\Delta x_{total} - \Delta x_{plates}}{\frac{\Delta T}{q} - \frac{\Delta x_{plates}}{k_m}}$$

where $q$ is applied heat flux along the geometry boundary, $\Delta x_{total}$ and $\Delta x_{plates}$ are the thicknesses of the entire geometry and of the plates (Figure 2), $\Delta T$ is the final steady-state temperature differential across the entire domain (including the end plates), $\Delta x_{total}$ is the total distance between the outer surfaces of the two plates, $\Delta x_{plates}$ is the combined thickness of the two plates, and $k_m$ is the material thermal conductivity used in the simulation, which is the same for the plates as it is for the spheres. All adjacent boundaries are insulated such that the net heat flux occurs only in one direction.

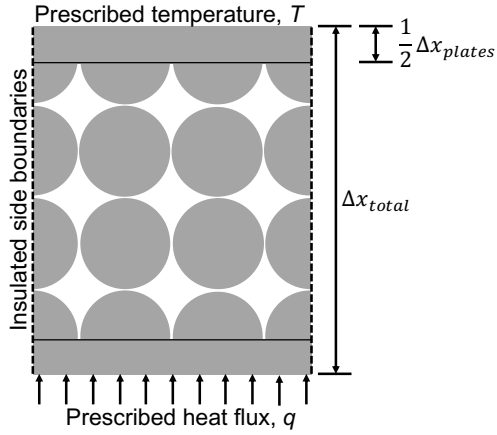

**Figure 2.** Cross-sectional illustration of packed sphere and end plate geometry that is used for steady-state thermal conductivity determination (Equation 4). Plate thicknesses are exaggerated and are not to scale. See Figure 3 for an example 3D geometry.

Multiple simulations for a single geometry are typically conducted across a range of temperatures. In this work, the fixed-temperature plates were typically set to temperatures of 250, 300, 350, and 400 K. The prescribed flux values ($q$) were chosen based on the estimated bulk thermal conductivity of the geometry so that a thermal gradient ($\Delta T$) on the order of 15 K would be achieved at steady state. The resulting thermal conductivity that is determined at the end of the simulation is assumed to be representative of the mean temperature of the geometry at the final steady state. If the prescribed heat flux is too high, the thermal gradient at steady state would be very large, which makes it dubious to use the mean temperature to represent the bulk conductivity given the temperature dependence of radiative heat transfer. If the flux is too small, the final temperature gradient value might be so small that precision is lost.

Heat conduction calculations for the tetrahedral domain mesh elements within the three-dimensional model geometry are performed using an implicit diffusion equation solver that is part of the CIMLib scientific FE library (Digonnet et al., 2007). Radiative heat transfer between triangular surface mesh elements is calculated following the Stefan-Boltzmann law in the following manner for element $i$ of $n$ elements:

(Eq. 5)
$$Q_i = \sum_{j=1}^{n} A_i F_{ij} \sigma (T_i^4 - T_j^4)$$

where $A$ is the area, $F_{ij}$ is the view factor from element $i$ to $j$ (i.e. the fraction of radiation leaving surface $i$ that strikes surface $j$), and $\sigma$ is the Stefan-Boltzmann constant. This value is added explicitly to the heat diffusion calculations, such that the surface temperatures from the previous time step are used for the calculation in Equation 5. A good discretization of the surface of the particles requires a large number of triangular surface elements. Thus, the number of possible pairs of surface elements that could "see" each other is high and requires an efficient computation of the view factors. The Intel Embree high-performance ray-tracing library meets these demands by rapidly tracing rays between all surface elements that share direct line of sight. View factors for all non-obstructed elements are calculated using the single line integration method described by Mitalas & Stephenson (1966) (c.f. Walton, 1986). The surface mesh size is refined such that the precise calculation of partially obstructed view factors between elements is not necessary, given that these types of calculations tend to be slower and less accurate (Walton, 2002). Emissivity is assumed to be unity, and as such, multiple reflections are not calculated. Our view factor calculations and explicit radiative heat transfer solver were validated with a geometric configuration with a known analytical solution — two concentric, hollow spheres are assigned fixed-temperature boundary conditions at the inner surface of the inner sphere and at the outer surface of the outer sphere (see Appendix A). The resulting steady-state temperatures of the outside of the inner sphere and the inside of the outer sphere were found to be within 1.5% of the analytical solution. The analytical expression for this test case is derived in Appendix A.

The geometries used in this study consist of three-dimensional beds of spheres that have been cropped to the shape of a parallelepiped box. Sphere coordinates are generated with the optimized dropping and rolling method described by Hitti & Bernacki (2013), following user-specified SFDs. We limit the range in particle sizes within a given geometry to a factor of ~6 to avoid the need for extremely large and computationally expensive representative volume elements. Plates are added to the top and bottom of the volume to create uniform surfaces for the heat flux and fixed temperature boundary conditions. The geometries are rendered and meshed using the Netgen automatic three-dimensional tetrahedral mesh generator (Schöberl, 1997) (Figures 3 and 4).

We performed a size convergence study with a simple cubic packing structure to determine the optimal representative volume element (i.e. parallelepiped width and height) where boundary effects are minimized. Model widths of only four particle diameters produced homogenized thermal conductivity results that were within ~5% of the convergence value for a much wider volume. Similarly, a model thickness of four to five sphere diameters produced thermal homogenized conductivity results within ~5% of the convergence value. As such, we conducted all further analyses with model geometries widths and heights of at least four sphere diameters. Commonly, we used larger geometries for random packings to capture a larger, more representative sampling of the packing. Most simulations thus contained between 120 and 175 spheres. Typical element counts were 500,000 to 3 million for volumetric (tetrahedral) elements and 130,000 to 500,000 surface (triangular) elements. View factor computation times ranged from a few minutes up to about 6 hours. Convergence in each heat flow simulation was typically achieved within 3–24 hours, depending on the number of elements and the number of processors used.

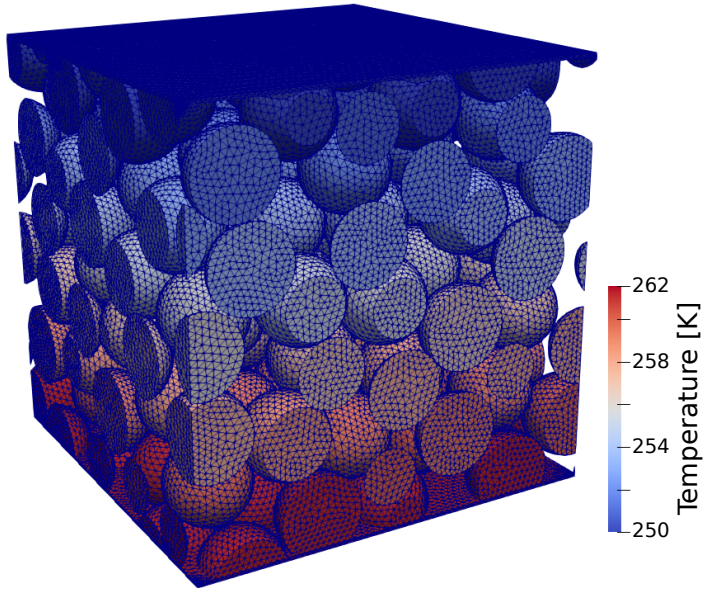

**Figure 3.** Meshed monodisperse sphere geometry colorized with final steady-state temperatures from a numerical simulation where a constant heat flux was applied to the bottom boundary plate and a fixed temperature (250 K) was imposed on the top plate. In this particular case, the spheres are not touching, so heat transfer between spheres is accomplished solely by radiation. In other conditions, as explained in the text and shown in Figure 4, we generated contact bridges between spheres to study the combined effect of solid and radiative heat transfer in a particulate medium.

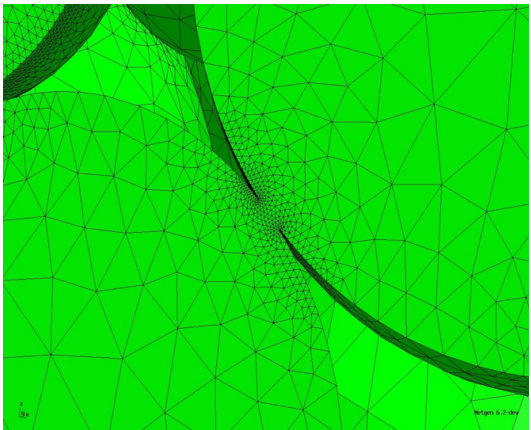

**Figure 4.** Example mesh refinement in a case where the spheres are touching and the ratio of particle contact neck size $r_c$ to sphere radius $R_p$ = 0.05.

Contacts between spheres can be approximated by adding cylindrical bridges at each contact location with radii equal to the desired contact radius $r_c$. It is necessary to use a fine mesh in the vicinity of the sphere contacts, given the narrowness of the geometry and the expectation of steep thermal gradients due to thermal constriction resistance (Figure 4). Following the numerical work by De Beer (2014), we use a minimum mesh element size that is approximately $0.2*r_c$. To validate the solid conductivity within our model, we performed a series of tests with ordered packing structures where the size of the contact neck, $r_c$, was systematically varied

(Figure 5). Our numerical results are in very good agreement with theoretical predictions by Siu & Lee (2000). Simulations with sphere contacts are generally more computationally intensive than those where the spheres are not in contact owing to the increased number of mesh elements. Therefore, although we have demonstrated the utility of this model for studying solid conduction through particulates via sphere-to-sphere contacts, the remainder of this work focuses on heat transfer between non-touching spheres by radiation only.

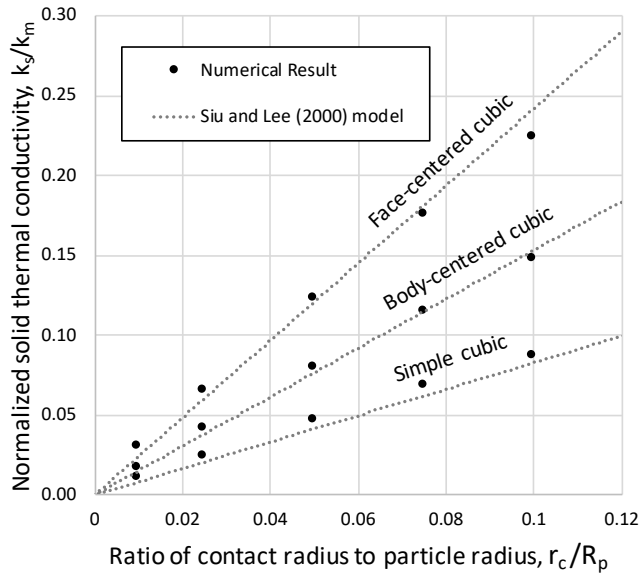

**Figure 5**. Solid conductivity $k_s$ (no radiation) normalized to the conductivity of the sphere material $k_m$ in regular particulate packings as a function of the ratio of $r_c$ to $R_p$.

To study the thermal conductivity of particulates due to radiation only, we reduced the diameter of the spheres by 0.1–0.5% to avoid overlaps. This was beneficial for computation times, as the removal of the contacts between spheres greatly reduced the mesh element count. We performed a sensitivity analysis to determine whether these slight reductions in sphere radius would appreciably affect the radiative thermal conductivity. We found that 1% radius reduction lowered the thermal conductivity by <1%. The applied change to sphere radius is thus assumed to be negligible.

In Figure 6, we compare the radiative thermal conductivity of a dense random packing of two test cases with monodisperse spheres with 2-mm and 1-cm diameters to theoretical model predictions. The model by Sakatani et al. (2017) predicts the radiative thermal conductivity results of the numerical simulations with an accuracy of approximately ±5%. Although in these cases their model overpredicted our results for smaller spheres and underpredicted our model for larger spheres, results from other test cases showed no systematic trend with particle diameter. We conclude from these tests that the Sakatani et al. (2017) model can be used to reliably predict the radiative thermal conductivity of a monodisperse sample, which we use to our advantage to compare polydisperse geometries to monodisperse geometries with equivalent radiative thermal conductivity values.

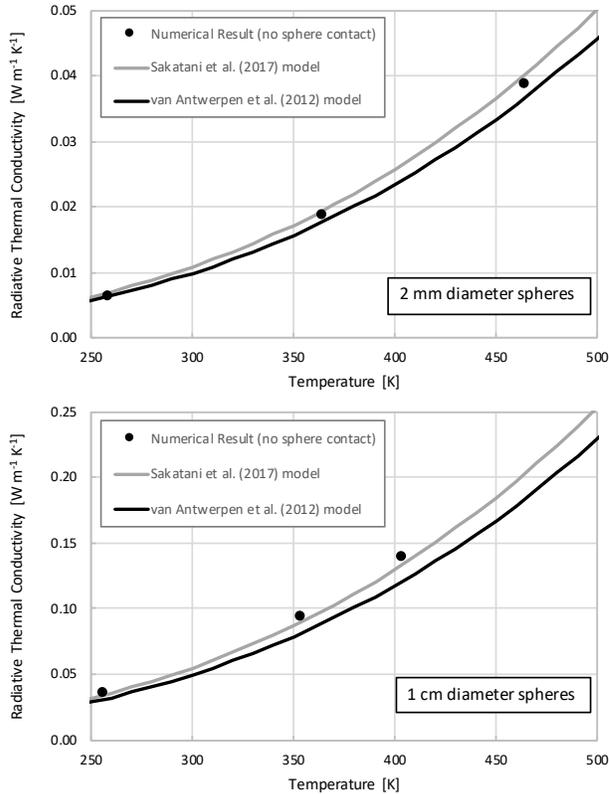

**Figure 6.** Radiative thermal conductivity results from numerical simulations of two monodisperse particulate geometries with diameters of 2 mm (upper plot) and 1 cm (lower plot, also shown in Figure 3). These results are compared to theoretical model predicted values. Temperatures were taken as the mean temperature from the numerical simulations.

## 4 Result for Monodisperse and Polydisperse Particulates

To study the radiative thermal conductivity ($k_r$) of particle size mixtures, we conducted numerical simulations of steady-state heat flow following the methodology described above where the spheres are not touching each other. We analyzed 16 randomly packed polydisperse particulate geometries with particle SFDs following a cumulative power law (Table 1). We used five different millimeter- to centimeter-scale particle size ranges, some of which were directly based on the size ranges described in Bierhaus et al. (2018) for tests of the OSIRIS-REx Touch-and-Go Sample-Acquisition Mechanism (TAGSAM) tests. These size ranges are relatively narrow due to the possible absence of fines on Itokawa (Bierhaus et al., 2018), which may also true for Bennu and Ryugu. Also, wider size ranges would require larger representative volumes, which could prohibitively increase the number of spheres and thus mesh elements. We generated packings for each size range by using various different cumulative power-law exponent values (e.g., Figure 7). We performed steady-state heat flow simulations at four different temperatures (approximately 250, 300, 350, and 400 K) to determine the radiative thermal conductivity as a function of temperature. We conducted all simulations with a solid material thermal conductivity value $k_m$ of 1.0 W m$^{-1}$ K$^{-1}$. Many of the simulations were recomputed with higher and lower $k_m$ values of 10.0, 0.5, 0.1, and 0.05 W m$^{-1}$ K$^{-1}$ to study the non-isothermality effect (Equations 1 and 2).

**Table 1.** Particle diameter size ranges used for polydisperse particulate models and the power-law exponent values used for each size range. The second and third row size ranges correspond to the "7c" and "7d" size distributions used by Bierhaus et al. (2018) for OSIRIS-REx TAGSAM sample collection testing. The geometries given in the last row (10–30 cm diameter) are shown in Figure 7.

| Particle diameter (mm) | | |
|---|---|---|
| min | max | Power law exponent values |
| 3.18 | 12.7 | –2.5, –3.0, –3.5 |
| 3.18 | 15.88 | –2.5, –3.0, –3.5 |
| 3.18 | 19.05 | –2.5, –3.0, –3.5 |
| 6.35 | 19.05 | –2.5, –3.0, –3.5 |
| 10.0 | 30.0 | –3.0, –4.0, –6.0, –8.0 |

An example set of results for one packing geometry as a function of temperature and $k_m$ is presented in Figure 8. In the simulations where $k_m$ is high ($\geq 1.0$ W m$^{-1}$ K$^{-1}$), the radiative thermal conductivity scales with the cube of the temperature, as predicted by theory. Lower values of $k_m$ cause the radiative thermal conductivity to decrease, particularly for higher temperatures. In extreme cases (e.g., $k_m = 0.05$ W m$^{-1}$ K$^{-1}$), the radiative conductivity was found to increase nearly linearly with temperature, rather than with temperature to the third power. These observations are similar to the non-isothermality described by van Antwerpen et al. (2012). Non-isothermality may be quantified as the ratio between the numerical radiative thermal conductivity results with non-isothermality ($1/\Lambda_s < 0.04$) and results without non-isothermality ($1/\Lambda_s > 0.04$). For example, we calculated non-isothermality correction factor ($f_k$) values for the numerical radiative thermal conductivity results in Figure 8 by calculating the ratio between all points on the plot and the solid theoretical curve. The calculated values of $f_k$ from all polydisperse simulations and our monodisperse validation simulations are plotted in Figure 9 against $1/\Lambda_s$ (Equation 1). In the case of the polydisperse geometries, we used the effective particle diameters (described in the next paragraph) to calculate the values of $\Lambda_s$ in Figure 9.

It is illustrative to express the conductivity results for these polydisperse geometries in terms of a monodisperse particle diameter that has the same thermal conductivity. We found that, when non-isothermality is negligible (i.e. $1/\Lambda_s < 0.04$), the radiative thermal conductivity of each polydisperse geometry is equivalent to the radiative thermal conductivity of a monodisperse geometry with some *effective* particle diameter. To determine the effective particle diameter for each polydisperse geometry, we used the Sakatani et al. (2017) model to calculate predicted monodisperse radiative thermal conductivity values. The model input diameter was allowed to vary as a free input parameter, whereas porosity and average coordination number were known from the polydisperse geometries. The determination of effective particle diameter was done only with the numerical thermal conductivity simulation solutions that we expected to be unaffected by non-isothermality—i.e., $1/\Lambda_s$ was less than ~0.04. We expect the uncertainty in the determined effective particle sizes to be approximately ±5%, on the basis of the goodness of fit between the Sakatani model predictions and our monodisperse numerical test cases. The resulting effective particle diameters are compared to volumetric mean particle size, volumetric median particle size, and the surface-volume mean particle diameter (Sauter mean) of each polydisperse geometry in Figure 10. The Sauter mean diameter, $D_{32}$, was calculated as the average of the volume-to-surface ratio of $n$ particles by:

(Eq. 6)
$$D_{32} = \frac{\sum_{i=1}^{n} D_i^3}{\sum_{i=1}^{n} D_i^2}$$

where $D_i$ is the diameter of the $i$th particle.

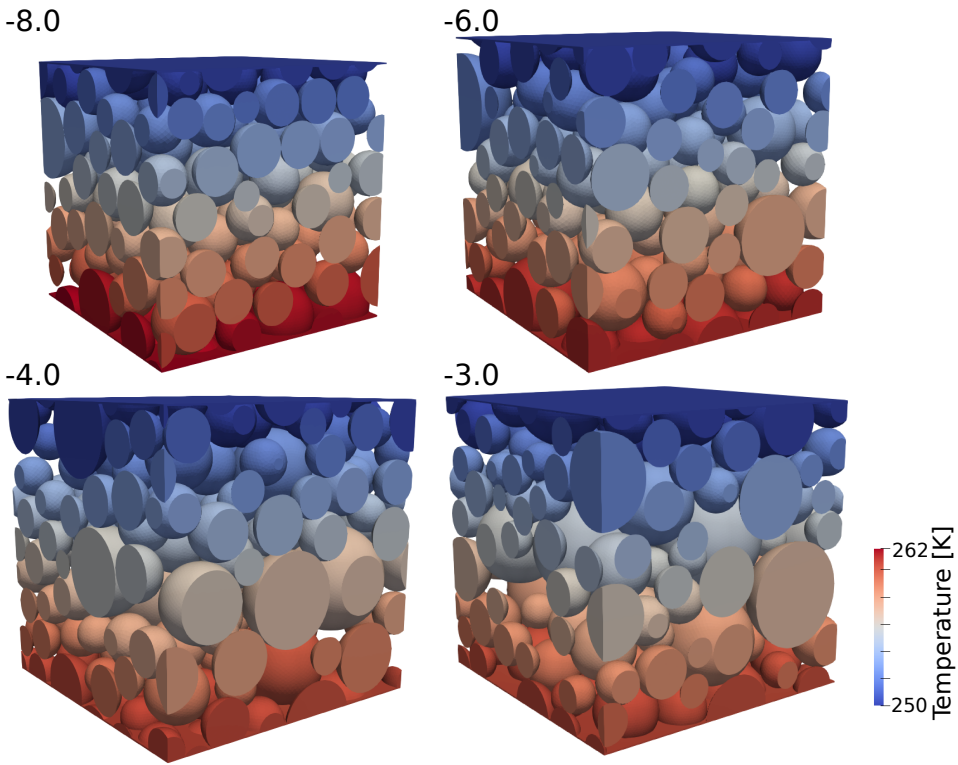

**Figure 7.** Steady-state temperature solution examples for the 10–30 mm diameter particle size range with four different power-law exponent values: –8.0, –6.0, –4.0, and –3.0. The material solid thermal conductivity used in these simulations is $k_m = 10.0$ W m$^{-1}$ K$^{-1}$, chosen to minimize non-isothermality within the individual spheres.

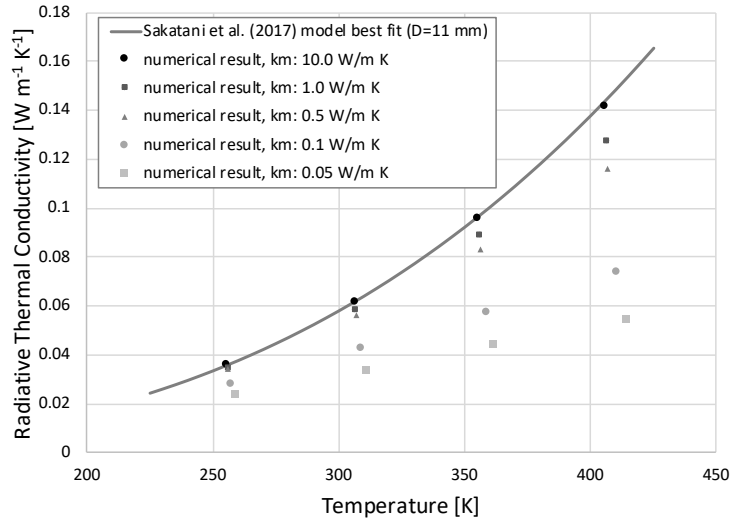

**Figure 8.** Numerical radiative thermal conductivity results for a mixture of particles having diameters 6.35–19.05 mm with an exponential slope of –2.5 as a function of temperature and solid particle material thermal conductivity $k_m$. The effective particle diameter for the case where $k_m = 10.0$ W m$^{-1}$ K$^{-1}$ is well fit by the theoretical model of Sakatani et al. (2017) with an effective particle diameter of 11 mm. It is clear that lowering the value of $k_m$ can substantially reduce the radiative thermal conductivity of the particulates, particularly at higher temperatures. Mean temperature from the numerical simulations were used to plot the x-axis temperatures here. For most simulations, the range in temperatures ($\Delta T$) between the two plates at steady state was less than 15 K. For simulations with very low values of $k_m$, $\Delta T$ was as high as 30 K.

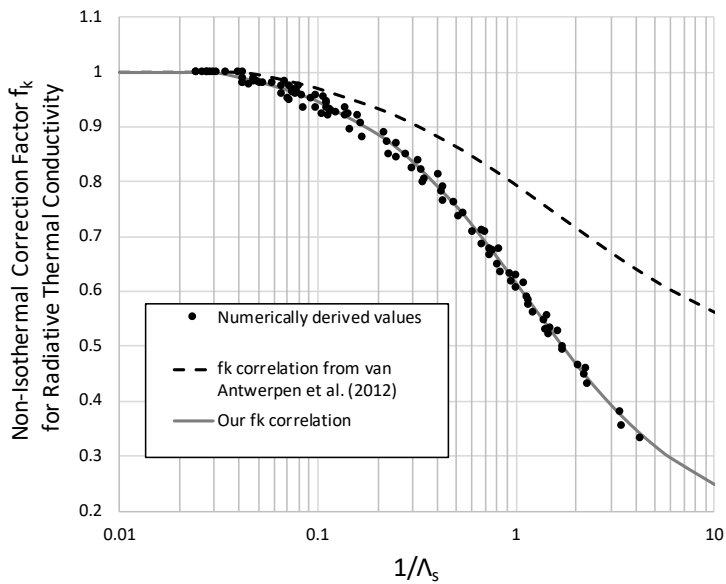

**Figure 9.** Original non-isothermal correction factor proposed by van Antwerpen et al. (2012) (dashed line), shown with our values for the correction factor from numerical results (points) and a corresponding fit line (solid gray line). $\Lambda_s$ values were calculated using the effective particle diameter for each geometry. Both lines are calculated using Equation 2, assuming that $f_k = 1$ when the equation produces a value greater than unity (when $1/\Lambda_s < {\sim}0.04$). Numerical test

results are shown for all polydisperse geometries in Table 1 and a set of monodisperse validation cases (e.g. Figures 3 and 6) at temperatures ranging from 250–400 K and with $k_m$ values ranging from 0.05–10.0 W m$^{-1}$ K$^{-1}$.

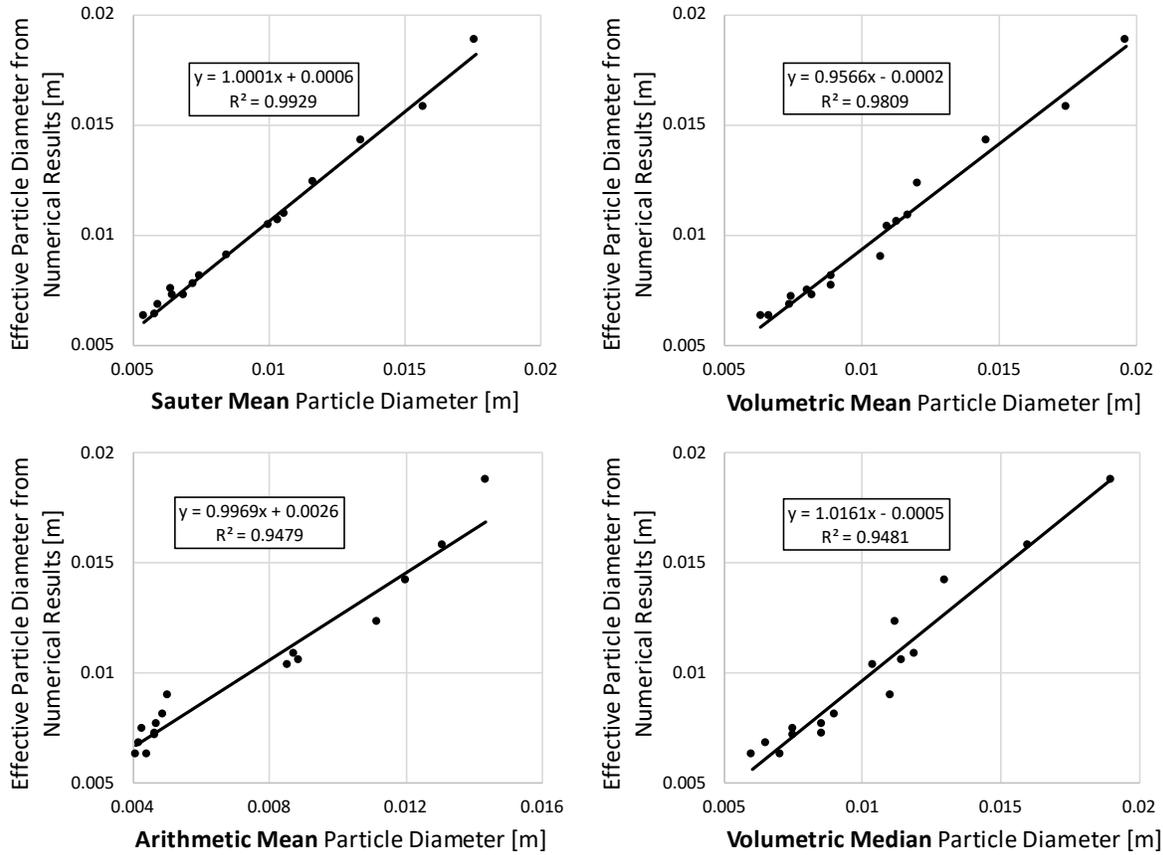

**Figure 10.** Comparison of the effective particle diameter determined from the radiative thermal conductivity numerical results to different expressions of the mean particle diameter for the 16 polydisperse geometries (Table 1). Effective particle diameter of a polydisperse particulate geometry is defined as the diameter of a monodisperse particulate geometry that has the same radiative thermal conductivity.

## 5 Discussion

The new values of $f_k$ versus $1/\Lambda_s$ in Figure 9 provide compelling evidence that an isothermal correction factor correlation exists that is applicable to both monodisperse and polydisperse particulates. We find that the $f_k$ values are well fit by Equations 2 and 3 using the follow coefficient values:

$a_1 = -0.568$
$a_2 = 0.912$
$a_3 = 0.765$
$a_4 = 1.035$

The value of $f_k$ approaches unity as $1/\Lambda_s$ drops to 0.04. Any value of $f_k$ is assumed to be unity when $1/\Lambda_s < \sim 0.04$, as shown in Figure 9. The original correction factor correlation proposed by van Antwerpen et al. (2012) underestimates the effects of non-isothermality that were observed in our simulations (Figure 9). As mentioned in Section 2, their correction factor values were determined from a numerical model of only short-range radiation between directly adjacent spheres and had not been validated for long-range radiation. Our correlation accurately accounts for both short- and long-range radiation together and appears to be valid under a wide range of particle SFDs. We are confident that our correlation is not simply recording a bias in our numerical results, given that it is composed of numerical solutions from geometries with multiple different length and width dimensions, temperature differentials, absolute temperatures, mesh sizes, and boundary flux values. However, all geometries tested here are dense random packings of spheres with unit emissivity, whereas the regolith on Bennu, Ryugu, and other airless bodies is surely composed of non-spherical particles, possibly with a much wider range of porosities and non-unit emissivity. It is difficult at this time to estimate what this porosity range might be, for example, given that particle angularity is expected to increase porosity but broad SFDs may counter this by decreasing porosity (e.g. pore-filling dust). Nonetheless, future studies should investigate how porosity, particle shape, and emissivity also affect regolith particle non-isothermality.

Our non-isothermality results have important implications for the interpretation of thermal inertia on Bennu and Ryugu. Initial results from OSIRIS-REx and Hayabusa2 indicate that the rocks on both asteroids may have very low thermal conductivities. Recently, Grott et al. (2019) determined that the boulder on the surface of Ryugu has a thermal conductivity in the range of 0.06–0.16 W m$^{-1}$ K$^{-1}$, based on measurements by the MARA radiometer aboard the Hayabusa2 MASCOT lander. If the particulates that compose the regolith in finer-particulate regions have similarly low thermal conductivity values, the non-isothermal correction factor described here would be necessary to correctly interpret the surface thermal inertia values. As an illustrative example, we performed a particle size inversion using a hypothetical regolith thermal inertia value of 200 J m$^{-2}$ K$^{-1}$ s$^{-1/2}$ for Bennu (Figure 11). This value falls near the low end of thermal inertia values so far determined for Bennu (Rozitis et al., 2019), which could be indicative in some cases of particulate regoliths. To avoid model bias, we used three different models to produce particle size predictions for this thermal inertia value, where the thermal conductivity of the individual particles ($k_m$) is allowed to vary. We assume that particle density must also vary with $k_m$ due to changes in microporosity within the regolith particles. Following the work by Grott et al. (2019), we use the average results from two models for meteorite thermal conductivity as a function of porosity to determine the regolith particle microporosity that would correspond to each x-axis value of $k_m$ (Henke et al., 2019; Flynn et al., 2019). Density is then calculated by assuming an average CM meteorite grain density (2960 kg m$^{-3}$, Flynn et al., 2018). The predicted particle sizes are shown using the three models with and without the non-isothermality correction (Equation 2) applied to the radiative thermal conductivity term. For comparison, the thermal conductivity of the Cold Bokkeveld CM2 chondrite meteorite (Opeil et al. 2010) and the boulder on Ryugu (Grott et al., 2019) are included as vertical line and a vertical, gray bar.

Finally, the diurnal skin depth (i.e. the approximate depth at which the amplitude of the thermal wave is attenuated by a factor of 1/e) for a 200 J m$^{-2}$ K$^{-1}$ s$^{-1/2}$ thermal inertia surface is also shown in Figure 11 as a thick gray line. The skin depth is calculated by:
(Eq. 7)
$$\delta = \frac{TI}{\rho c_p}\sqrt{\frac{P}{\pi}}$$

The rotation period, $P$, for Bennu is 15466 seconds (Lauretta et al., 2019), thermal inertia, $TI$, is assumed at 200 J m$^{-2}$ K$^{-1}$ s$^{-1/2}$, specific heat, $c_p$, is 750 J kg$^{-1}$ K$^{-1}$ (Biele et al., 2019), and bulk density, $\rho$, is estimated as a function of $k_m$, using the average of two models by Henke et al. (2016) and Flynn et al. (2018) (see Methods section in Grott et al., 2019), assuming a grain density of 2960 kg m$^{-3}$ (Flynn et al., 2018) and a regolith macroporosity of 0.40.

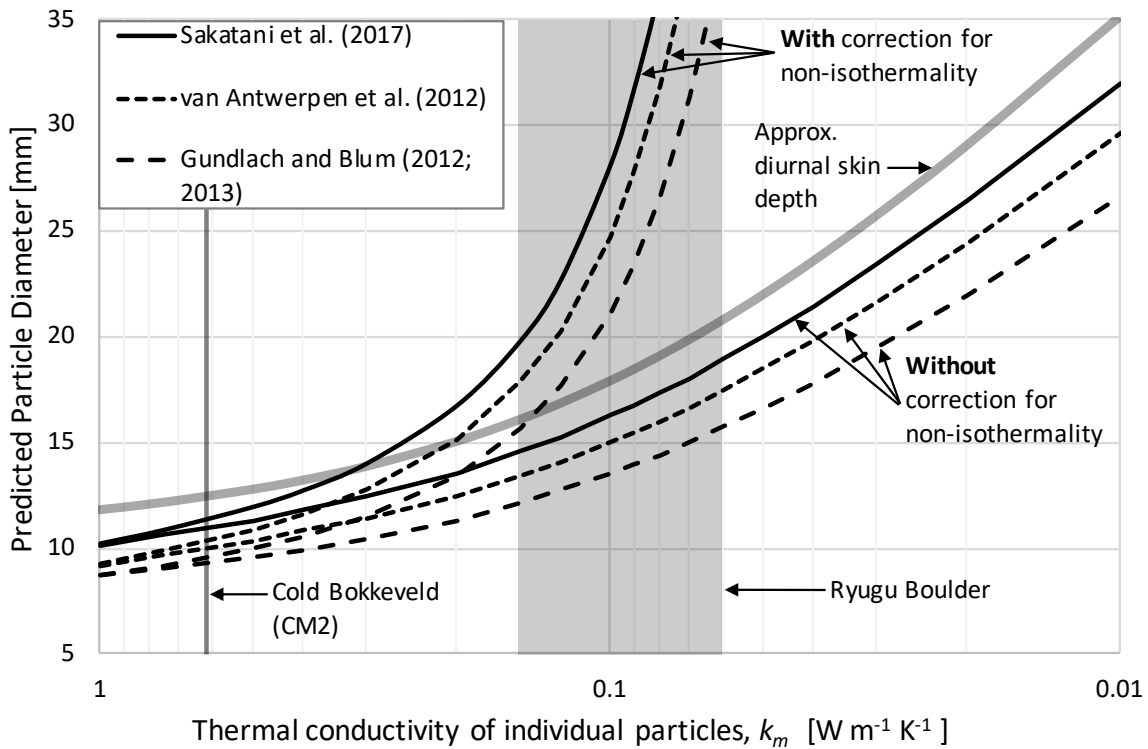

**Figure 11.** Example particle size prediction for a hypothetical surface on Bennu with a thermal inertia of 200 J m$^{-2}$ K$^{-1}$ s$^{-1/2}$ at 265 K as a function of the conductivity of the material that makes up the regolith particles ($k_m$). Three regolith thermal conductivity models are used, both without and with the use of the non-isothermal correction factor, to make particle size predictions. For simplicity, only one macroporosity (0.40) is considered here. Increasing the assumed porosity would have the effect of further increasing the predicted particle diameters. The predicted diurnal skin depth is shown as a thick, gray line. Two reference $k_m$ values are displayed with vertical lines/bars: The Cold Bokkeveld (CM2) meteorite (Opeil et al., 2010) and the Ryugu boulder examined by the MARA radiometer data (Grott et al., 2019). Model input values: macroporosity=0.40; specific heat =750 J kg$^{-1}$ K$^{-1}$ (a); particle density is a function of thermal conductivity, using the average of two models (b,c) for meteorite thermal conductivity vs microporosity and assuming an average CM grain density=2960 kg m$^{-3}$ (c); surface energy=0.032 J m$^{-2}$ (d); emissivity=1.0; Bennu mid-latitudes surface acceleration=40 μm s$^{-2}$ (e); Poisson's ratio=0.27 (f); Young's Modulus=5.63e9 Pa (f); mean burial depth=10 mm; Sakatani model tunable solid contact conductivity parameter ξ=0.63 (d); tunable radiative conductivity parameter ζ=0.85 (d). References for model inputs: [a]Biele et al. (2019), [b]Henke et al. (2016), [c]Flynn et al. (2018), [d]Sakatani et al., (2018), [e]Scheeres et al. (2019), [f]A.R. Hildebrand (personal communication, 2018), measured values for Tagish Lake simulant.

The importance of considering non-isothermality is immediately apparent in Figure 11, where it is shown that the assumed value of $k_m$ can affect the particle size predictions by a factor of ~2–3. However, we must consider the gray diurnal skin depth line in the figure, which can be considered an approximate upper cutoff for the validity of the particle size determination. This is based on the well-known idea that regoliths composed of particles that are much larger than the diurnal skin depth are expected to behave thermally like solid rock (e.g. Christensen, 1986). Particle size predictions that exceed the diurnal skin depth should be considered indeterminate, such that the particles could be *any* size larger than the approximate diurnal skin depth. Interestingly, we see in Figure 11 that if the non-isothermality effect is ignored, one might conclude that a thermal inertia of 200 J m$^{-2}$ K$^{-1}$ s$^{-1/2}$ is consistent with effective particle diameters in the range of ~10–25 mm, depending on assumptions of $k_m$, density, and macroporosity. However, once non-isothermality is accounted for, the predicted particle size increases markedly as $k_m$ is reduced. If the particles have a $k_m$ value that is similar to the Ryugu boulder from Grott et al. (2019) (grey shaded vertical bar), then one can only conclude that the regolith is predominantly composed of particles larger than ~15 mm. It very well could just be boulders or a bedrock. Conversely, if we assume that the regolith particles have a $k_m$ that is equivalent to the Cold Bokkeveld meteorite, one could conclude that this thermal inertia observation is consistent with a regolith with an effective particle size near 10 mm.

In addition to our findings related to non-isothermality, we have also found that a polydisperse particulate generally has a radiative thermal conductivity that is equivalent to a monodisperse particulates with some effective particle diameter (Figure 8, solid line), at least in the case of particulates that follow cumulative power law SFDs. We also found that the effective particle diameter can be used in Equation 1 to calculate $\Lambda_s$ for the determination of the non-isothermality correction factor, $f_k$, for a polydisperse regolith.

Figure 10 compares the effective particle diameters for 16 polydisperse geometries to various expressions for the mean and median particle diameter in a particle size mixture. The best correspondence, nearly 1:1, is found with the Sauter mean particle diameter. The Sauter mean diameter may be physically interpreted as the diameter that could be used to produce a monodisperse population of spheres that has the same surface area and the same volume (but not the same number of spheres) as the polydisperse sphere population (Kowalczuk & Drzymala,

2016). In the context of our work, this means that the radiative thermal conductivity of a polydisperse population of spheres is the same as the radiative thermal conductivity of a population of monodisperse spheres that has the same total mass (or volume) and same total surface area—suggesting that the total surface area present within a particulate volume is an important control on the radiative thermal conductivity. This is perhaps not surprising, given that heat diffusion through opaque particulates is largely limited by radiative heat transfer between the particle surfaces. If surface area is indeed a control, then replacing the spheres with any other particle shape would have the effect of increasing the total surface present within the system (if porosity is kept the same) and thus should decrease the radiative thermal conductivity. This idea is consistent with experimental datasets by Sakatani et al. (2017), in which they found that monodisperse, angular basaltic particles from the JSC-1A lunar simulant had a lower radiative thermal conductivity than spherical glass particles with equivalent diameters. Whether this is truly due to changes in surface area or instead to some other factor, such as changes in the length scales of the void spaces, is an interesting subject of investigation for a future study.

In Figure 10, there appears to also be a nearly 1:1 correspondence between effective particle diameter and volumetric median particle size. This correlation supports experimental results by Sakatani et al. (2018), who found that the thermal conductivity of the JSC-1 lunar regolith simulant was equivalent to the volumetric median particle size for both radiative and solid thermal conductivity terms. However, the particle diameters deviate from this trendline in Figure 10 by up to ±15%. As such, thermal inertia measurements of a regolith with an unknown SFD could be used to estimate the volumetric median particle size of the regolith to within about ±15% of the true value. One could also use thermal data to estimate the Sauter mean particle size with a higher degree of accuracy, but the volumetric median particle size is perhaps a more useful metric because of its more common use in statistical reports. We must however take caution when applying these results to terrains with very broad SFD. In such cases, the largest particles in the SFD could exceed the diurnal skin depth in size and may thus have thermophysical behavior that is distinct compared to the surround finer regolith. This configuration cannot be approximated in a thermal model as a single material with a single set of representative thermal properties, but instead would need to be approximated as a lateral mixture of materials with different thermal properties (e.g. a checkerboard mixture of boulders and regolith).

Although these findings related to regolith particle SFDs might aid us in interpreting regolith properties from thermal inertia, we must again point out the importance of the relationship between regolith particle size and diurnal skin depth, particularly for analyses on fast rotators with apparently low thermal conductivity boulders like Bennu and Ryugu. If the effective particle size in some region is larger than the diurnal skin depth, the apparent thermal inertia is expected to be that of the solid rock, thus precluding particle size determination from thermal data. As seen in Figure 11, even the lower end of the thermal inertia values on Bennu (200 J m$^{-2}$ K$^{-1}$ s$^{-1/2}$) corresponds to particle size predictions that approach or exceed this cutoff. Although this complicates or even prevents the interpretation of regolith thermal inertia in terms of particle size, it might present an opportunity to predict the thermal properties of the individual regolith particles: The OSIRIS-REx spacecraft will obtain very high-resolution photos as well as a series of coincident OTES and OVIRS spectra during the sampling event. Should the site contain abundant fine-particulate regolith, the regolith particle SFD, particle shape, and packing configuration could be constrained from these data, providing a valuable ground truth for thermal models. An analysis like the one shown in Figure 11 could be repeated with constraints on the effective particle size in order to estimate the value of $k_m$. If $k_m$ is low, then perhaps the regolith particles have a high degree of microporosity and are derived from the low density/conductivity boulders. Conversely, if $k_m$ is high—perhaps even higher than the surrounding boulders—then this could indicate that the regolith particles are stronger clasts weathered out of a weaker, lower density/conductivity boulder matrix. Either scenario would provide interesting information regarding the nature of Bennu's boulders and regolith formation processes.

## 6 Conclusions

In this work, we build upon the fundamental knowledge of particulate regolith thermal properties in an effort to improve our ability to infer the physical properties (namely particle size distribution) of regolith from remote thermal measurements. This work was motivated by the ongoing OSIRIS-REx mission at asteroid Bennu. Radiative thermal conductivity was chosen as the focus here owing to the dominance of the radiative heat transfer (relative to heat conduction through the contacts between particles) for coarse, centimeter-scale regolith, which is the material of interest for sampling on Bennu.

We presented a new thermophysical model for asteroid regolith where the individual regolith particles are represented as randomly packed spheres in a FE mesh framework. Heat conduction and surface-to-surface radiation are modeled in a steady-state configuration to determine the bulk conductivity properties of the particulates. With this model, we determined the radiative thermal conductivity of a series of polydisperse particle geometries with cumulative power-law SFDs. We found that the thermal conductivity of the particulate material can have a very strong influence on the radiative thermal conductivity, particularly when material thermal conductivity ($k_m$) is very low, the temperature is high, and/or particles are large, owing to non-isothermality within individual particles (Figure 8). When this non-isothermality is negligible, the radiative thermal conductivity of polydisperse geometries scales normally with the cube of the temperature, as is also the case for monodisperse particulates. When non-isothermality is substantive, the radiative thermal conductivity has a weaker temperature dependence. The magnitude of non-isothermality for polydisperse and monodisperse particulates is larger than previously predicted, and new correlation coefficients ($a_1 = -0.568$, $a_2 = 0.912$, $a_3 = 0.765$, $a_4 = 1.035$) were provided for a correction factor (Equation 2) for the radiative thermal conductivity. This correction factor can be applied to any model for regolith thermal conductivity (e.g. Sakatani et al., 2017; Gundlach and Blum, 2013) by multiplying the model-predicted radiative thermal conductivity term by $f_k$. If the regolith particles on Bennu have very low thermal conductivity, particle predictions from thermal inertia could be notably underestimated if non-isothermality is not taken into account in the regolith thermal conductivity model.

Finally, we found that the radiative thermal conductivity for each polydisperse simulation is identical to that of a monodisperse simulation with some effective particle diameter. A 1:1 correspondence exists between the effective particle diameter determined for the numerical results for each polydisperse geometry and the Sauter mean particle diameter (Figure 10). We conclude that amount of surface area per unit mass and volume within a particulate geometry is closely related to the total radiative conductivity, leading to the prediction that non-spherical particles should have lower radiative thermal conductivity values. The effective particle diameters of the mixtures also correspond within ±15% to the volumetric median particle diameter, enabling a convenient means of interpreting thermophysical data on the regolith of asteroid Bennu and other airless bodies, particularly when ancillary information about the regolith such as cumulative power-law slope and maximum particle size can be gained from image data.

**Acknowledgments and Data**

The authors acknowledge support from the Academies of Excellence on Complex Systems and Space, Environment, Risk and Resilience of the Initiative d'EXcellence (IDEX) Joint, Excellent, and Dynamic Initiative (JEDI) of the Université Côte d'Azur as well as from the Centre National d'Études Spatiales (CNES). This material is based in part upon work supported by NASA under Contract NNM10AA11C issued through the New Frontiers Program. We thank Christopher Edwards and an anonymous reviewer for providing useful comments that helped to improve the quality of this manuscript. We also thank Josh Emery, Jens Biele, Dante Lauretta, and Cat Wolner for useful suggestions and edits. Particulate geometry files and a spreadsheet of model input parameters and output values are available in an external archive (Ryan et al., 2019; http://doi.org/10.5281/zenodo.3575067).

**Appendix A**

For the radiation validation test, we define a steady-state configuration with two concentric spherical shells with surface emissivity values of unity. We number the four surfaces as 1–4, starting from the inner surface of the inner spherical shell (Figure A1). Fixed temperatures are defined at surfaces 1 and 4, while surfaces 2 and 3 exchange energy via radiation. Once final steady state is achieved, the final temperatures of surfaces 2 and 3 are related to the net energy transfer between surfaces 2 and 3 by:

$$Q = A_2 \sigma f_{2-3}(T_2^4 - T_3^4) \qquad (1)$$

where $Q$ is energy transfer in watts, $A_2$ is the area of surface 2, $f_{2-3}$ is the view factor or configuration factor for surface 2 to surface 3, which in this case is equal to one, $\sigma$ is the Stefan-Boltzmann constant, and $T$ is the temperature of the surface denoted by the subscript.

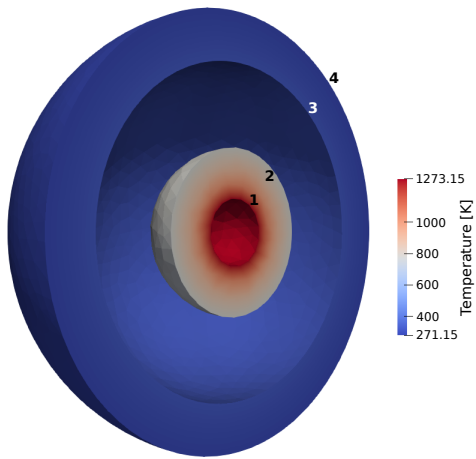

**Figure A1.** Hollow spheres radiation validation case. The four surfaces referenced in the text are labeled.

To describe steady-state heat conduction through the spherical shells, we start with the one-dimensional form of the heat equation in spherical coordinates:

$$\frac{1}{r^2}\frac{d}{dt}\left(r^2\frac{dT}{dr}\right) = 0 \qquad (2)$$

The first integral gives:

$$r^2\frac{dT}{dr} = C_a \qquad (3)$$

where $C_a$ is in integration constant. From Fourier's Law:

$$Q = -kA\frac{dT}{dr} \qquad (4)$$

we determine $C_a$:

$$C_a = \frac{-Qr^2}{kA} \qquad (5)$$

Taking the second integral of the heat equation (Equation 2) gives:

$$T(r) = -\frac{C_a}{r} + C_b \qquad (6)$$

For the case of the inner spherical shell, we can evaluate this equation at $r_1$ and $r_2$ to determine the value of $C_b$:

$$T_1 = \frac{-C_a}{r_1} + C_b$$
$$T_2 = \frac{-C_a}{r_2} + C_b$$
$$C_b = T_1 + \frac{C_a}{r_1} = T_2 + \frac{C_a}{r_2} \qquad (7)$$

We can substitute this into Equation 6 to find:
$$T(r) = T_1 + C_a\left(\frac{1}{r_1} - \frac{1}{r}\right) \qquad (8)$$

Evaluating at $r = r_2$ provides an expression for $T_2$:

$$T_2 = T_1 - \frac{Qr_2^2}{kA_2}\left(\frac{1}{r_1} - \frac{1}{r_2}\right) \qquad (9)$$

Similarly, an expression for $T_3$ is found to be:

$$T_3 = T_4 - \frac{Qr_3^2}{kA_3}\left(\frac{1}{r_4} - \frac{1}{r_3}\right) \qquad (10)$$

Substituting Equation 9 and 10 into Equation 1 gives us the final expression for net heat transfer in the system:

$$Q = A_2 \sigma \left[ \left( T_1 - \frac{Q r_2^2}{k A_2} \left( \frac{1}{r_1} - \frac{1}{r_2} \right) \right)^4 - \left( T_4 - \frac{Q r_3^2}{k A_3} \left( \frac{1}{r_4} - \frac{1}{r_3} \right) \right)^4 \right] \qquad (11)$$

Equation 11 may be solved for $Q$ numerically. In our work, we used the *fsolve()* function within the SciPy Python library. Once $Q$ is found, the values of $T_2$ and $T_3$ are found from Equations 9 and 10 and can be compared to the results of the numerical test when steady state is achieved.